\documentclass[authoryear,12pt,5p,times]{elsarticle}
\def\apj{ApJ}
\usepackage{times}
\usepackage{aas_macros}

\usepackage{amsmath}
\usepackage{amssymb}
\usepackage[utf8x]{inputenc}
\usepackage{graphicx}
\usepackage{epsfig}
\usepackage{latexsym}
\usepackage{color}
\usepackage{dcolumn}% Align table columns on decimal point
\usepackage{bm}% bold math
\usepackage{natbib}
\usepackage{setspace}
\usepackage{subfigure}
\usepackage{psfrag}
\usepackage{listings}
\usepackage{mathrsfs}
\usepackage{listings}
\usepackage{enumerate}
\usepackage[table]{xcolor}
\usepackage{hyperref}
\definecolor{Red}{rgb}{1, 0, 0}
\definecolor{Green}{rgb}{0, 1, 0}
\definecolor{Blue}{rgb}{0, 0, 1}
\definecolor{Black}{rgb}{0, 0, 0}
\definecolor{Grey}{rgb}{0.5, 0.5, 0.5}
\definecolor{White}{rgb}{1, 1, 1}
\definecolor{Yellow}{rgb}{1, 1, 0}
\definecolor{Magenta}{rgb}{1, 0, 1}
\definecolor{Cyan}{rgb}{0, 1, 1}
\definecolor{myCyan}{rgb}{0, 0.6, 1}
\definecolor{Orange}{rgb}{1, 0.5, 0}
\definecolor{Violet}{rgb}{0.5, 0, 0.5}
\definecolor{DarkRed}{rgb}{0.5, 0., 0.3}
\definecolor{Pink}{rgb}{1., 0.4, 0.7}
\definecolor{LightPink}{rgb}{1, 0.8, 0.8}
\definecolor{YellowGreen}{rgb}{0.6, 0.8, 0}
\definecolor{LightYellow}{rgb}{1., 1., 0.95}
\definecolor{Brown}{cmyk}{0, 0.8, 1, 0.6}

\journal{Astronomy and Computing}
\begin{document}
%
%%%%%%%%%%%%%%%%%%%%%%%%%%%%%%%%%%%%%%%%%%%%%%%%%%%%%%%%%%%%%%%%%%%%%%%%%%%%%%%%
%
\begin{frontmatter}

\title{AMADA-Analysis   of Multidimensional  Astronomical Datasets}


\author[MTA]{R S. de Souza}\ead{rafael@caesar.elte.hu}
\author[MPA]{B.~Ciardi}
\author{for the COIN collaboration}

\address[MTA]{MTA E\"otv\"os University, EIRSA ``Lendulet'' Astrophysics Research Group, Budapest 1117, Hungary}
\address[MPA]{Max-Planck-Institut f\"ur Astrophysik, Karl-Schwarzschild-Str. 1, D-85748 Garching, Germany}

\begin{abstract}

We present AMADA,  an interactive  web application to  analyse  multidimensional datasets.   The user uploads a simple   \textsc{ascii} file  and  AMADA performs  a number of exploratory  analysis together   with  contemporary  visualizations diagnostics. 
The package performs a  hierarchical clustering in the parameter  space, and  the user can choose among  linear, monotonic  or   non-linear correlation analysis. AMADA provides a number of  clustering visualization  diagnostics such as heatmaps, dendrograms,  chord diagrams, and graphs.  In addition,  AMADA has the option to run  a standard or robust principal components analysis, displaying the results as polar bar plots.    
The code is  written  in {\sc r} and the web interface  was created using the  {\sc shiny} framework. AMADA source-code is freely available at \href{https://goo.gl/KeSPue}{https://goo.gl/KeSPue}, and the shiny-app at  \href{http://goo.gl/UTnU7I}{http://goo.gl/UTnU7I}. 
 \end{abstract}

\end{frontmatter}

%
%________________________________________________________________
% / --------------------
\topmargin -1.3cm
% NOTE THAT THIS WILL PLACE THE LAYOUT CORRECTLY ON ASTRO-PH
% PLEASE LEAVE IN
% -------------------- /

%%%%%%%%%%%%%%%%%%%%%%%%%%%%%%%%%%%%%%%%%%%%%%%%%%%%%%%%%%%%%%%%%%%%%%%%%%%%%%%%%%%%%%%%%%%%%%%%%%%%%%%%%%%%%%
\section{Introduction}

The emerging precision era of astronomy marks the transition from a data-deprived field to a data-driven science, in which statistical methods play a central role. The need to handle these ever-increasing  datasets impacts all branches of modern science, characterizing the so-called era of Big Data.  As a consequence, an efficient exploration   of  high-dimensional datasets is becoming ubiquitous throughout all  scientific fields, such as biology   \citep[e.g.,][]{Venter2004}, social sciences \citep[e.g.,][]{Patty2015}, geology \citep[e.g.,][]{Terence2014} and astronomy  \citep[e.g.,][]{Ball2010, Graham2013, Martinez2013}.

Upcoming surveys such as  the Large Synoptic Survey Telescope \citep[e.g.,][]{Anderson2009}, the Square Kilometre Array \citep[e.g.,][]{Carilli2014}, and Euclid \citep[e.g.,][]{Scaramella2015},  just to mention a few,    will push the boundaries of our ability to analyse sky catalogs, while the ever-increasing  complexity of    cosmological simulations  keeps lessening   the distance between observed and synthetic data \citep[e.g., ][]{Overzier2013, deSouza2013b,deSouza2014b,Vogelsberger2014}. 
  
 An optimal exploration of these catalogs, observed and/or simulated, heavily relies on our ability to uncover hidden relationships among different quantities \citep[e.g.,][]{Borne2008,Ball2010,Graham2013},  such as fundamental planes of galaxy properties \citep{Tully1977,Faber1976}, as well as to identify the optimal set of variables to describe and predict a certain property of interest (e.g. the presence of star formation activity in a halo; \citealt{deSouza2015}).

A mainstay   methodology  for   data exploration in astronomy is  the correlation analysis.  Its goal is  to describe the level of association, usually linear, between a given pair of variables. Its  applicability virtually covers the entire  astronomical domain,  such as   gamma-ray bursts \citep[e.g.,][]{Burgess2014}, cosmic voids \citep{Hamaus2014},  star formation activity \citep{Lee2013}, dark matter halo properties \citep{deSouza2013a,deSouza2014a}, and   baryonic galaxy properties  \citep{Yates2012}, just to cite a few.

To facilitate the use of contemporary exploratory and visualization techniques commonly used in other scientific fields but not fully exploited in astronomy, we developed the AMADA package.
The code allows the user to visualize subgroups of variables with high  association in a hierarchical tree structure through diverse visual tools, such as graphs, chord diagrams, dendrograms and heatmaps. The goal is to deliver  a user-friendly guide for a first data screening. By providing a systematic methodology for clustering detection in the space of object properties, the researcher can make a statistically justified decision about the subset of  features to be studied in a given catalog.

It is worth noting that  other  interfaces for data exploration in astronomy exist \citep[e.g,][]{Brescia2010,Burger2013,Konstantopoulos2015}. Particularly,   VOStat \citep{Chakraborty2013} and AstroStat \citep{Kembhavi2015} are two web-based services for statistical analysis using {\sc R}  under the hood.  Both projects are  focused on providing a user-friendly environment to  perform a wide range of  standard statistical analysis, such as hypothesis testing, multivariate analysis, clustering and so forth. However,  AMADA is the first of its kind with a primary focus on  information visualization techniques for   general correlation  analysis in  multidimensional catalogs.

%%%%%%%%%%%%%%%%%%%%%%%%%%%%%%%%%%%%%%%%%%%%%%%%%%%%%%%%%%%%%%%%%%%%%%%%%%%%%%%%%%%%%%%%%%%%%%%%%%%%%%%%%%%%%%

%%%%%%%%%%%%%%%%%%%%%%%%%%%%%%%%%%%%%%%%%%%%%%%%%%%%%%%%%%%%%%%%%%%%%%%%%%%%%%%%%%%%%%%%%%%%%%%%%%%%%%%%%%%%%%
\section{Main features}

AMADA is written in \textsc{r} 3.1.1 and developed using Rstudio\footnote{\url{www.rstudio.com}} and Shiny\footnote{\url{shiny.rstudio.com}} frameworks.  RStudio is an open source interface for development of \textsc{r} applications, and  Shiny is a  package that allows to  build interactive web applications directly  from \textsc{r}. Instructions on how to run the code locally,  and a brief installation tutorial are given in  \ref{app:shiny}. 

The package allows an interactive exploration and information retrieval from  high-dimensional datasets. The user can choose among  different methods for correlation analysis, whose outcomes are displayed in a chosen graphical layout for  visual inspection.  In the following, we briefly describe the main available  features.

\subsection{Datasets}

\begin{figure*}
\centering
\includegraphics[trim= 0.2cm 2cm 0.2cm 4cm, clip=true, width=2\columnwidth]{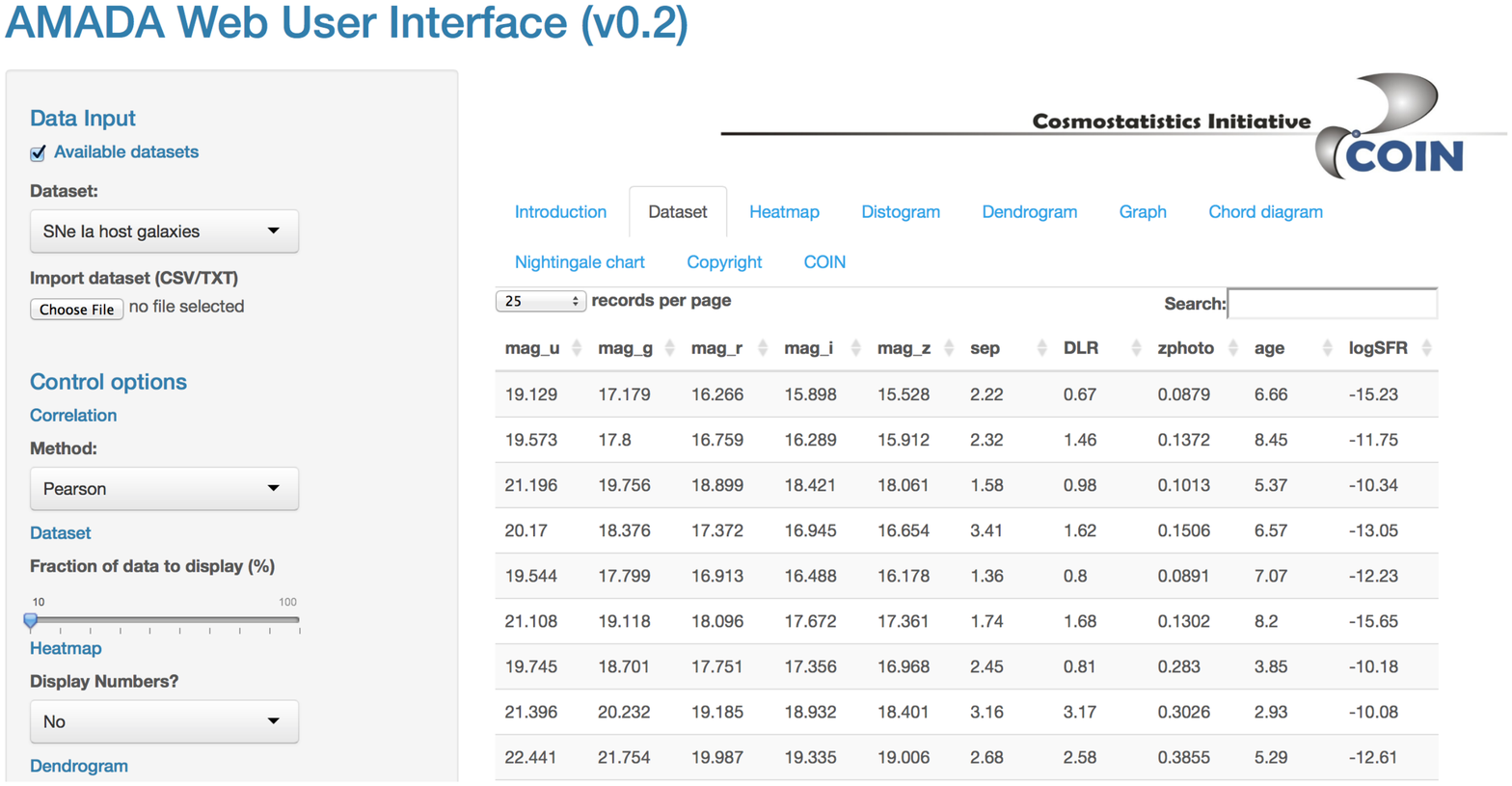}
\caption{A screenshot of the AMADA portal showing properties of host  galaxies of Type Ia supernovae. This portal is publicly available at \href{http://goo.gl/UTnU7I}{http://goo.gl/UTnU7I}.}
\label{fig:dataset}
\end{figure*}

The user can  upload a dataset in a plain text \textsc{ascii} file as space or comma separated values (CSV).  The columns should be named, and missing data should be marked as $NA$. An example of how a typical dataset  looks like,  together with a screenshot  from the web portal, is displayed in Fig.  \ref{fig:dataset}.  Alternatively, the user  can  use the  \textit{download data} button to  inspect on its own text editor how to  format the matrix. The current version of AMADA does not allow an interactive selection of columns. Therefore,  we show below how it can be easily done in  {\sc r} command line using the $c$ function:

\begin{lstlisting} 
data(iris)
colnames(iris)<-c("SL","SW","PL","PW","Species")
head(iris)
    SL   SW   PL   PW   Species
    5.1  3.5  1.4  0.2   setosa
    4.9  3.0  1.4  0.2   setosa
iris2<-iris[,c("SL","SW")]
head(iris2)
    SL   SW
    5.1  3.5
    4.9  3.0
\end{lstlisting}
The original column names of the famous iris dataset \citep{FISHER1936} are shortened in the example (S = sepal, P = petal, L = length, W = width) to save space.

In addition, some public catalogs are already made available on the portal. In the following we will use two of them for explanatory purposes. As an example of low-dimensional and relatively small sample we use a catalog of galaxies experiencing  supernova (SN) explosions, while as an example of high-dimensional and moderately large sample we use a mock galaxy catalog. More specifically, we apply AMADA to investigate:

\begin{itemize}

\item Supernova host galaxy properties \citep[][]{Sako2014}. In this catalog the properties of Type Ia and II supernova host galaxies are retrieved from the Sloan Digital Sky Survey multi-band photometry. The available catalog represents a sub-sample of the original one, after removal of non-supernova objects and missing data. The final sample is composed of 443 (56) galaxies hosting Type Ia (Type II) supernova, each of them described by 10 parameters, such as galaxy age, star formation rate, distance from supernova to the host galaxy, and so forth.

\item Galaxy properties \citep{Guo2011}. A mock galaxy catalog built  using semi-analytic galaxy formation models and the N-body Millennium Simulations \citep{Springel2005}. 
 The initial   data set   is composed of $\approx 180,000 $   haloes  at  redshift  0.   To avoid numerical artifacts due to low resolution effects, we select only those structures   with at least $300$ particles \citep[e.g.,][]{Antonuccio2010}. In addition, we consider only central  star forming  galaxies (i.e., no satellite galaxies).  The remaining  dataset  is  composed of 7079  haloes, and each halo is described by approximately  30 parameters.
 \end{itemize}
As here we adopt the original nomenclature for the various quantities, we recommend the reader to refer to the original articles or catalogs for a detailed description of each parameter.  
%%%%%%%%%%%%%%%%%%%%%%%%%%%%%%%%%%%%%%%%%%%%%%%%%%%%%%%%%%%%%%%%%%%%%%%%%%%%%%%%%%%%%%%%%%%%%%%%%%%%%%%%%%%%%%

%%%%%%%%%%%%%%%%%%%%%%%%%%%%%%%%%%%%%%%%%%%%%%%%%%%%%%%%%%%%%%%%%%%%%%%%%%%%%%%%%%%%%%%%%%%%%%%%%%%%%%%%%%%%%%%%%%%%%%%%%%%%%%%%%%%%%%
\subsection{Control Options}

Several control options are available on the portal to choose among different methods of analysis and visualization. Once the desired combination is chosen, the user should click on the button \textit{Make it so!} to update the results. The following options are available:

\begin{itemize}

\item Fraction of data to display: choose the percentage of data displayed on the screen.

\item Correlation method: choose among Pearson, Spearman or Maximum Information Coefficient (MIC).

\item Display numbers: choose if correlation coefficients should be displayed in the heatmap.

\item Dendrogram type: choose among phylogram, cladogram or fan configurations\footnote{Visualizations  inspired by  phylogenetic tools  \citep[e.g.,][]{Paradis2004}.}.

\item Graph layout: choose between spring and circular configurations.

\item Chord diagram colour: choose among different colour schemes.

\item Number of PCs: choose the number or Principal Components (PCs) to display as Nightingale charts.

\item PCA method: choose between  standard  or robust Principal Components Analysis (PCA).

\end{itemize}
%%%%%%%%%%%%%%%%%%%%%%%%%%%%%%%%%%%%%%%%%%%%%%%%%%%%%%%%%%%%%%%%%%%%%%%%%%%%%%%%%%%%%%%%%%%%%%%%%%%%%%%%%%%%%%

%%%%%%%%%%%%%%%%%%%%%%%%%%%%%%%%%%%%%%%%%%%%%%%%%%%%%%%%%%%%%%%%%%%%%%%%%%%%%%%%%%%%%%%%%%%%%%%%%%%%%%%%%%%%%%%%%%%%%%%%%%%%%%%%%%%%%%
\section{Methods}

In this section we briefly discuss the different methods used by AMADA to analyse the datasets.

\subsection{Correlation methods }

The correlation analysis quantifies  the strength of the association between a pair of  variables, through a correlation coefficient. Its absolute  value varies between 0 (uncorrelated variables) and 1 (perfect association). Currently, AMADA offers three options of correlation measurements:  linear \citep[Pearson;][]{Pearson1895}, monotonic \citep[Spearman;][]{Spearman1904} and non-linear \citep[MIC;][]{Reshef2011}. We briefly present them in the following, and refer the reader to the original papers for more details.

%%%%%%%%%%%%%%%%%%%%%%%%%%%%%%%%%%%%%%%%%%%%%%%%%%%%%%%%%%%%%%%%%%%%%%%%%%%%%%%%%%%%%%%%%%%%%%%%%%%%%%%%%%%%%%
\paragraph*{Pearson}

This is widely employed in statistics to measure the degree of the relationship between linearly related variables.  The following formula is used to estimate the Pearson coefficient,  $r_p$,  between two variables $X_i$ and $Y_i$:

\begin{equation}
r_p = \frac{\sum_{i=1}^n(X_i-\overline{X})(Y_i-\overline{Y})}{\sqrt{\sum_{i=1}^n(X_i-\overline{X}^2)}\sqrt{\sum_{i=1}^n(Y_i-\overline{Y})^2}}, 
\end{equation}
where $\overline{X}$ and $\overline{Y}$ represent the sample mean, and $n$ the total number of objects in the dataset.  

\paragraph*{Spearman rank correlation}
 
This is a non-parametric  method to measure the degree of monotonic association between two variables, and does not rely on any distributional assumption.   For a dataset of size $n$, the variables  $X_i$ and $Y_i$ are converted to ranks\footnote{In statistics, {\it ranking} refers to the data transformation in which numerical or ordinal values are replaced by their rank when the data are sorted. For example, if the numerical data 3.8, 5.4, 2.1, 10.3 are observed, the ranks of these data items would be 2, 3, 1 and 4 respectively.}, and the following formula is used to calculate the Spearman coefficient,  $\rho$:

\begin{equation}
\rho = 1 - \frac{6\sum_{i=1}^n d_i^2}{n(n^2-1)},
\end{equation}
where $d_i = R_{X_i}-R_{Y_i} $ is the difference between ranks. 
%%%%%%%%%%%%%%%%%%%%%%%%%%%%%%%%%%%%%%%%%%%%%%%%%%%%%%%%%%%%%%%%%%%%%%%%%%%%%%%%%%%%%%%%%%%%%%%%%%%%%%%%%%%%%%

%%%%%%%%%%%%%%%%%%%%%%%%%%%%%%%%%%%%%%%%%%%%%%%%%%%%%%%%%%%%%%%%%%%%%%%%%%%%%%%%%%%%%%%%%%%%%%%%%%%%%%%%%%%%%%
\paragraph*{Maximal information coefficient}

MIC \citep{Reshef2011} is founded under concepts of information theory \citep[e.g.,][]{Wentian1990}. In this context, the Shannon entropy,  $\mathcal{H}$,  can be understood as a measure of uncertainty of a random variable. For a  single discrete distribution it can be written   as 
\begin{equation}
\mathcal{H}(A) = -\sum_{a \in A}p(a)\log p(a), 
\end{equation}
while the joint entropy for a pair of discrete random variables ($A$,$B$) with a joint distribution $p(a,b)$ is defined as
\begin{equation}
\mathcal{H}(A,B) = -\sum_{a \in A}\sum_{b \in B}p(a,b)\log p(a,b), 
\end{equation}
where $p(a)$ and $p(b)$ are the marginal  probability mass  functions (PMFs)
of $A$ and $B$, and $p(a,b)$ is the joint PMF.
Hence, the mutual information ($\mathrm{MI}$)  measures the amount of information that one random variable contains about another random variable, 
\begin{eqnarray}
\mathrm{MI}(A, B) &=& \sum_{a \in A} \sum_{b \in B} p(a,b) \log \left( \frac{p(a,b)}{p(a)p(b)} \right),\nonumber\\
&\equiv& \mathcal{H}(A) - \mathcal{H}(A,B).
\end{eqnarray}
\noindent

Consider $D$ as a finite set of ordered pairs, $\{(a_i, b_i), i = 1, \ldots, n\}$,  partitioned into a $x$-by-$y$ grid of variable size, $G$, such that there are  $x$-bins spanning $a$ and $y$-bins covering $b$, respectively.
The PMF of a particular grid cell is  proportional to the number of data points  inside that cell.
We can  define a  characteristic matrix $M(D)$ of a set $D$ as
\begin{equation}
\label{eq:MD}
\mathrm{M(D)}_{x,y} =  \frac{\max(\mathrm{MI})}{\log \min \{x,y\}},
\end{equation}
representing the highest normalized $\mathrm{MI}$ of $D$.
The MIC of a set $D$ is then defined as
\begin{equation}
\mathrm{MIC(D)} = \max_{0 < xy < B(n)} \left\{\mathrm{M(D)}_{x,y} \right\}, 
\end{equation}
representing the maximum value of $M$ subject to $0 < xy <  B(n)$, where the function $B(n) \equiv n^{0.6}$ was empirically determined by \cite{Reshef2011}.
%%%%%%%%%%%%%%%%%%%%%%%%%%%%%%%%%%%%%%%%%%%%%%%%%%%%%%%%%%%%%%%%%%%%%%%%%%%%%%%%%%%%%%%%%%%%%%%%%%%%%%%%%%%%%%%%%%%%%%%%%%%%%%%%%%%%

%%%%%%%%%%%%%%%%%%%%%%%%%%%%%%%%%%%%%%%%%%%%%%%%%%%%%%%%%%%%%%%%%%%%%%%%%%%%%%%%%%%%%%%%%%%%%%%%%%%%%%%%%%%%%%%%%%%%%%%%%%%%%%%%%%%%
\subsection{Principal Components Analysis}

The ultimate goal of PCA is to reduce the dimensionality of a multivariate dataset,
while explaining the data variance  with as few PCs as possible.
Given its versatility, it  has been applied to a broad range of astronomical studies, such as 
stellar, galaxy and quasar spectra \citep[e.g.,][]{Chen2009, McGurk2010}, 
galaxy properties \citep{Conselice2006, Scarlata2007}, 
Hubble parameter and cosmic star formation reconstruction \citep[e.g.,][]{ishida2011a, ishida2011b}, and
supernova photometric classification \citep{ishida2013}.

PCA belongs to a class of Projection-Pursuit \citep[PP; e.g.,][]{Croux2007}  methods, whose aim is to detect structures in multidimensional data  by projecting them onto a lower dimensional subspace (LDS). The LDS is selected by maximizing a projection index (PI), where PI represents a given feature  in the data (trends, clusters, hyper-surfaces, anomalies, etc.).
The particular case where  variance ($S^2$) is taken  as a PI leads to the  classical version of PCA\footnote{
  The PCs are computed by diagonalization of the data covariance matrix ($\Sigma^2$),  with the resulting eigenvectors corresponding to PCs and the resulting eigenvalues to the variance {\it explained} by the PCs.
The eigenvector corresponding to the largest eigenvalue gives the direction of greatest variance (PC1), the second largest eigenvalue gives the direction of the next highest variance (PC2), and so on.
Since covariance matrices are symmetric positive semidefinite, the eigenbasis is orthonormal (spectral theorem).
}. The  PCA scheme employed here falls into the category of filter methods of feature selection. Their aim is to determine how relevant   is  a feature  in representing a class in a high-dimensional space, but there exist other approaches, i.e. the wrapper methods, that can be  tailored to determine how relevant a feature is  against a given classification task \citep[see e.g., ][for a discussion of feature selection methods in astronomy]{Donalek2013}.

Given $n$ parameters 
$x_1, \cdots, x_n$, all of them column vectors of dimension $\Gamma$, the first PC is obtained by finding a unit vector $\mathbf{a}$ which maximizes the variance of the data projected onto it:
\begin{equation}
\mathbf{a_1} =\underset{||\mathbf{a}||=1}{\rm \arg\max}~ S^2 (\mathbf{a}^tx_1,\cdots,\mathbf{a}^tx_n), 
\label{eq:PC1}
\end{equation}
where $t$ is the transpose operation and   $\mathbf{a_1}$ is the direction of the first PC\footnote{
  $\underset{x}{\arg\max}~  f(x)$ is the set of values of $x$ for which the function $f(x)$ attains its largest value.
}.
Once we have computed the $(k-1)$th PC, the direction of the $k$th component, for $1 < k \leqslant \Gamma$, is given by 
\begin{equation}
\mathbf{\mathbf{a}_k} = \underset{||\mathbf{a}||=1,\mathbf{a}\bot \mathbf{a}_1,\cdots,\mathbf{a}\bot \mathbf{a}_{k-1}}{\rm \arg\max}S^2(\mathbf{a}^tx_1,\cdots,\mathbf{a}^tx_n), 
\end{equation}
where the condition of each PC to be orthogonal to all previous ones ensures a new uncorrelated basis.
Despite of these attractive properties, the classical version of PCA has some critical drawbacks, as the  sensitivity to outliers  \citep[e.g.,][]{Hampel2005}.
In order to overcome this limitation,  several  robust versions were created. For instance, instead of taking the variance as a PI in equation (\ref{eq:PC1}), a robust
measure of variance \citep{Hoaglin2000} is taken, i.e.  
the median absolute deviation \citep[MAD; e.g., ][]{Howell2005} of an ordered set  $\kappa$ is given by 
\begin{equation}
\mathrm{MAD}(\kappa_1, \cdots,\kappa_n) = 1.48 \underset{j}{\mathrm{med}}|(\kappa_j - \underset{i}{\mathrm{med}}(\kappa_i)|), 
\end{equation}
where med represents the median of the sample, and the square of MAD gives the  robust  variance. The value of 1.48 represents $Q_{0.75}^{-1}$, where $Q_{0.75}$ is the 0.75 quantile of a normal distribution.  AMADA allows the user to run  a  robust PCA based on the grid  search base algorithm   from \citet{Croux2007}.
%%%%%%%%%%%%%%%%%%%%%%%%%%%%%%%%%%%%%%%%%%%%%%%%%%%%%%%%%%%%%%%%%%%%%%%%%%%%%%%%%%%%%%%%%%%%%%%%%%%%%%%%%%%%%%

%%%%%%%%%%%%%%%%%%%%%%%%%%%%%%%%%%%%%%%%%%%%%%%%%%%%%%%%%%%%%%%%%%%%%%%%%%%%%%%%%%%%%%%%%%%%%%%%%%%%%%%%%%%%%%
\subsection{Hierarchical Clustering}

A cluster analysis can be understood  as a descriptive statistics to determine if a given dataset  should be divided into   different  groups.  The method aims to identify which groups of  objects are similar to each other but different (or distant) from objects in other groups.  
There are several ways to define  dissimilarity (or distance), according to each particular goal. Since we are interested in finding  groups of variables highly correlated, it is natural to define the dissimilarity, $\mathcal{D}$, between properties  as 
\begin{equation}
\mathcal{D}(X_i,Y_i) = 1-\left|{\rm Corr}(X_i,Y_i)\right|, 
\end{equation}
where $\rm{Corr}$ stands for  correlation measurement.  
Thus, $\mathcal{D}(X_i,Y_i) = 0$ represents  perfect correlation, while the value of  $\mathcal{D}(X_i,Y_i) = 1$ indicates uncorrelated variables. 

One of the main advantages of hierarchical clustering methods is that a prior specification of the number of clusters to be searched is not needed. Instead, the method requires a measurement of dissimilarity between groups of variables, which is  based
on the pairwise dissimilarities among the observations 
within each of  two groups. 
We employ an  agglomerative approach, where  each variable is initially   assigned to its own cluster, then the method   recursively merges a selected pair of clusters into a single one, where each  new pair is  composed by  merging  the two
groups with the smallest  $\mathcal{D}$ in the immediately  lower level  of the hierarchy.  
The lowest level represents each single variable, while the   highest level is a single cluster  containing all variables. The final  outcome is a  hierarchical representations in which
the clusters at each level of the hierarchy are created by merging clusters at the next lower level. 
To guide the user in the task of selecting a certain sub-group of interest, we provide an optimal number of clusters estimated via the \citeauthor{Calinski74} index \citep{Calinski74}. 
The tree-like final structure can be  graphically portrayed by e.g.,  dendrograms, graphs and chord diagrams,  as discussed in the following \S \ref{sec:infovis}.
%%%%%%%%%%%%%%%%%%%%%%%%%%%%%%%%%%%%%%%%%%%%%%%%%%%%%%%%%%%%%%%%%%%%%%%%%%%%%%%%%%%%%%%%%%%%%%%%%%%%%%%%%%%%%%%%%%%%%%%%%%%%%%%%%%%%

%%%%%%%%%%%%%%%%%%%%%%%%%%%%%%%%%%%%%%%%%%%%%%%%%%%%%%%%%%%%%%%%%%%%%%%%%%%%%%%%%%%%%%%%%%%%%%%%%%%%%%%%%%%%%%%%%%%%%%%%%%%%%%%%%%%%
\section{Visualization tools}
\label{sec:infovis}

When dealing with a  large amount of complex information, visualizing it in an intelligible way becomes a challenge.  In this case, the aim of a visualization method  is 
to optimize the intuitive insight of the data structure  in order to  exploit the perceptual capabilities of the human eye. Whilst the role of visualization belongs to the groundwork
 of astronomical analysis, new paradigms for multidimensional data visualization are not fully  exploited,  when  compared to other fields. 
Patterns, trends and correlations that might go undetected in tabular-based data, can be revealed and more easily communicated with interactive  visualization tools.
AMADA incorporates  contemporary methods   to visualize multidimensional data properties and their intrinsic correlations. This is particularly relevant if one aims to have a physical intuition of possible sub-populations  of highly correlated quantities, which are not necessarily the dominant components of the whole sample.  In the following, we describe the main visual  capabilities of the package with a brief introduction of each methodology. 
%%%%%%%%%%%%%%%%%%%%%%%%%%%%%%%%%%%%%%%%%%%%%%%%%%%%%%%%%%%%%%%%%%%%%%%%%%%%%%%%%%%%%%%%%%%%%%%%%%%%%%%%%%%%%%%%%%%%%%%%%%%%%%%%%%%%

%%%%%%%%%%%%%%%%%%%%%%%%%%%%%%%%%%%%%%%%%%%%%%%%%%%%%%%%%%%%%%%%%%%%%%%%%%%%%%%%%%%%%%%%%%%%%%%%%%%%%%%%%%%%%%%%%%%%%%%%%%%%%%%%%%%%
\subsection{Heatmap}

The cluster heatmap is a rectangular grid representation of a matrix with cluster trees appended to its margins. Its aim is to  facilitate inspection of  cluster structures in large matrices within a compact displayed area.  The method is  broadly used  in the biological sciences \citep{Wilkinson2009}, and it is worth to  cite its recent  application to   solar data mining \citep[Fig. 10 of][]{Schuh2015}. 

\begin{figure*}
\centering
\includegraphics[width=2\columnwidth]{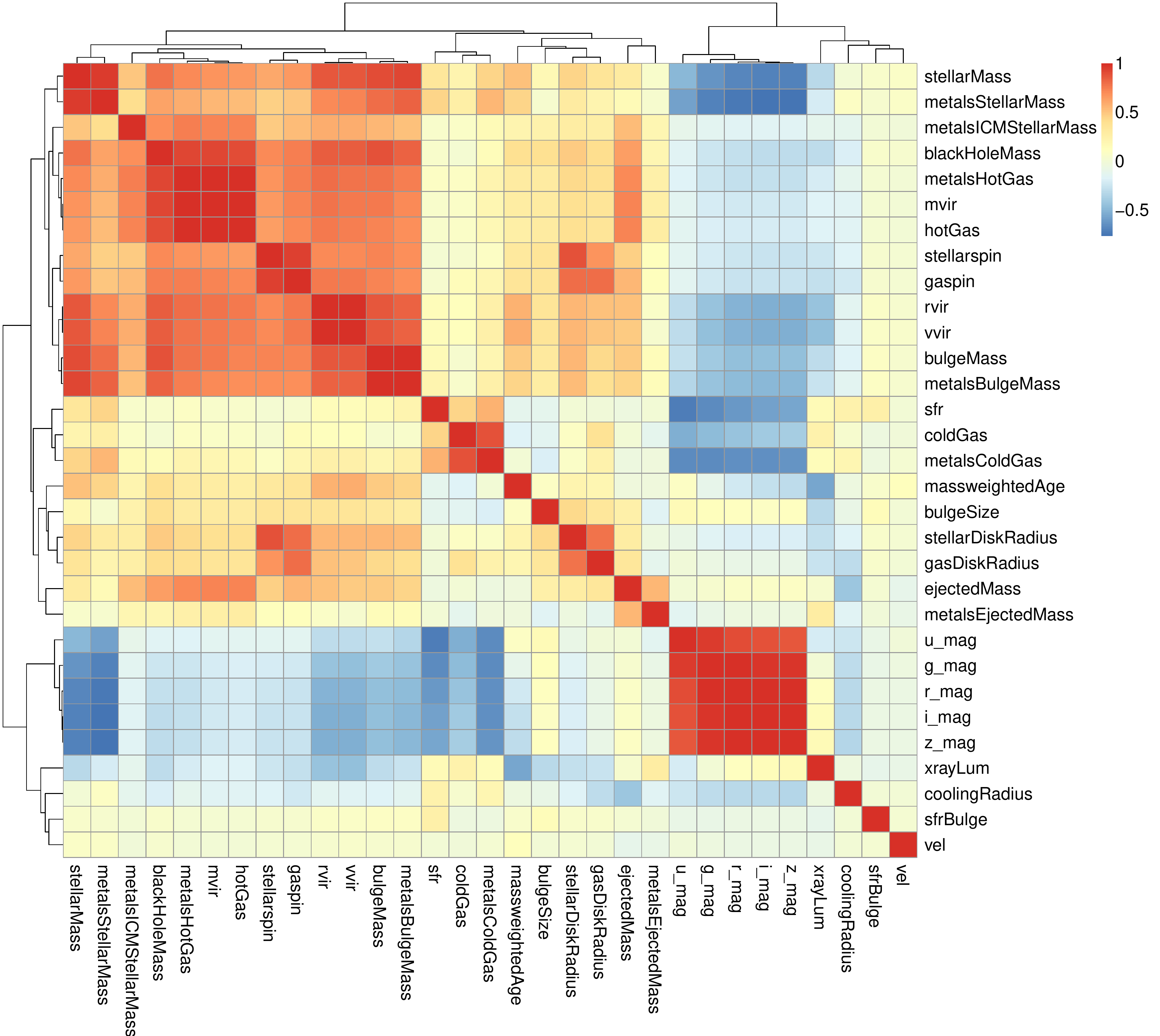}
\caption{Heatmap visualization of the correlation matrix (using a  Pearson correlation measure) of some galaxy properties from the mock galaxy catalog by \cite{Guo2011}.  Red indicates strong positive correlation and blue indicates strong negative correlation. Yellows are associated to correlations close to zero. 
  }
\label{fig:heatmap}
\end{figure*}

In case of a correlation matrix,  the color assigned to a point in the heatmap grid indicates how much each pair of variables correlates, as can be seen in the typical heatmap  shown in 
Fig. \ref{fig:heatmap}. 
For visualization purposes, the arrangement of the rows and columns is made following a hierarchical clustering with a dendrogram drawn at the edges of the matrix.  The figure portrays  the heatmap of the   mock galaxy catalog from \cite{Guo2011}. 
Note the red square in the bottom right corner of the panel, automatically highlighting the trivial association between the magnitudes in the $u, g, r, i$,  and $z$ bands. 
Less trivial  associations can be identified more easily using for instance  a dendrogram visualization,   as discussed in the following section. 

%%%%%%%%%%%%%%%%%%%%%%%%%%%%%%%%%%%%%%%%%%%%%%%%%%%%%%%%%%%%%%%%%%%%%%%%%%%%%%%%%%%%%%%%%%%%%%%%%%%%%%%%%%%%%%%%%%%%%%%%%%%%%%%%%%%%%%%%%%%%%%%%%%%%%%%%%%
\subsection{Dendrogram}
\label{sec:dend}

A dendrogram provides a comprehensive  description of
the hierarchical structures  in a visual  format. Among the applications in astronomical research   are the   hierarchical  structural analysis of  interstellar properties \citep{Houlahan1992},   molecular clouds \citep{Rosolowsky2008}, and explanatory classification of galaxies \citep{Fraix2012}.
The individual variables are arranged along the bottom of the dendrogram and referred to as leaf nodes.  Clusters are formed by joining individual variables  or existing  clusters, with the joint point referred to as a node.  At each dendrogram node we have a right and left sub-branch of clustered variables.  The height of the node can be understood  as the  dissimilarity $\mathcal{D}$ 
between the right and left sub-branch clusters.

\begin{figure}
\centering
\includegraphics[width=1.1\columnwidth]{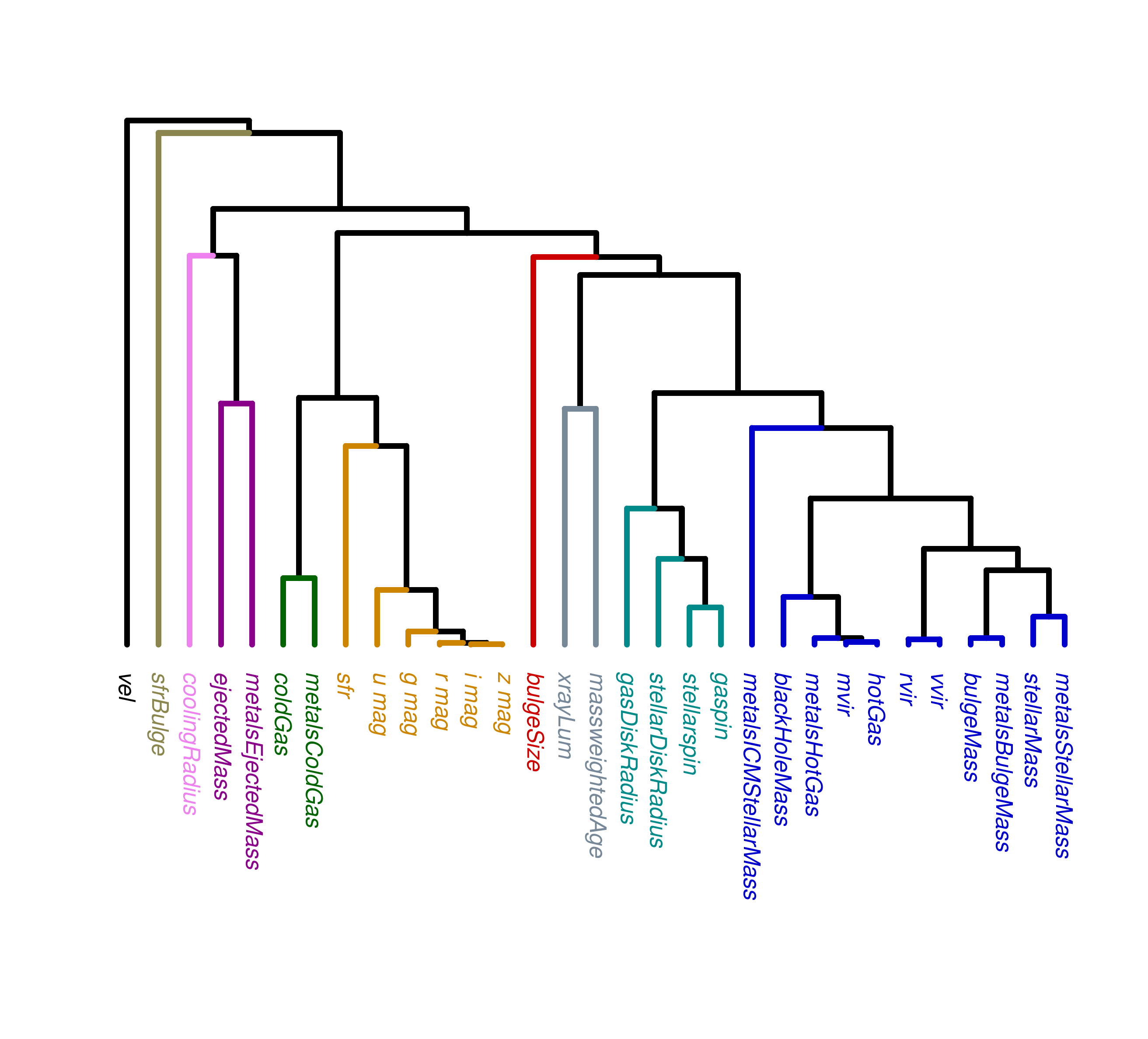}
\caption{Dendrogram of the galaxy properties from the \citet{Guo2011} catalog. The  different sub-groups  of galaxy properties, assigned using the \citeauthor{Calinski74} index,  are    colored according to the  cluster assignment. 
  }
\label{fig:dendrogram}
\end{figure}

Fig. \ref{fig:dendrogram}  displays a dendrogram of the galaxy properties from the \cite{Guo2011} catalog, divided in 10 major clusters (indicated by different colors) using the \citeauthor{Calinski74} index. 
The method automatically suggests interesting associations among the galaxy properties, such as the $u$-band as an indicator of the star formation rate \citep[SFR; see e.g.][]{Gilbank2010}.

%%%%%%%%%%%%%%%%%%%%%%%%%%%%%%%%%%%%%%%%%%%%%%%%%%%%%%%%%%%%%%%%%%%%%%%%%%%%%%%%%%%%%%%%%%%%%%%%%%%%%%%%%%%%%%%%%%%%%%%%%%%%%%%%%%%%
\subsection{Graphs}

\begin{figure*}
\centering
\includegraphics[trim= 0.2cm 0.5cm 0.2cm 0.5cm, clip=true,width=1\columnwidth]{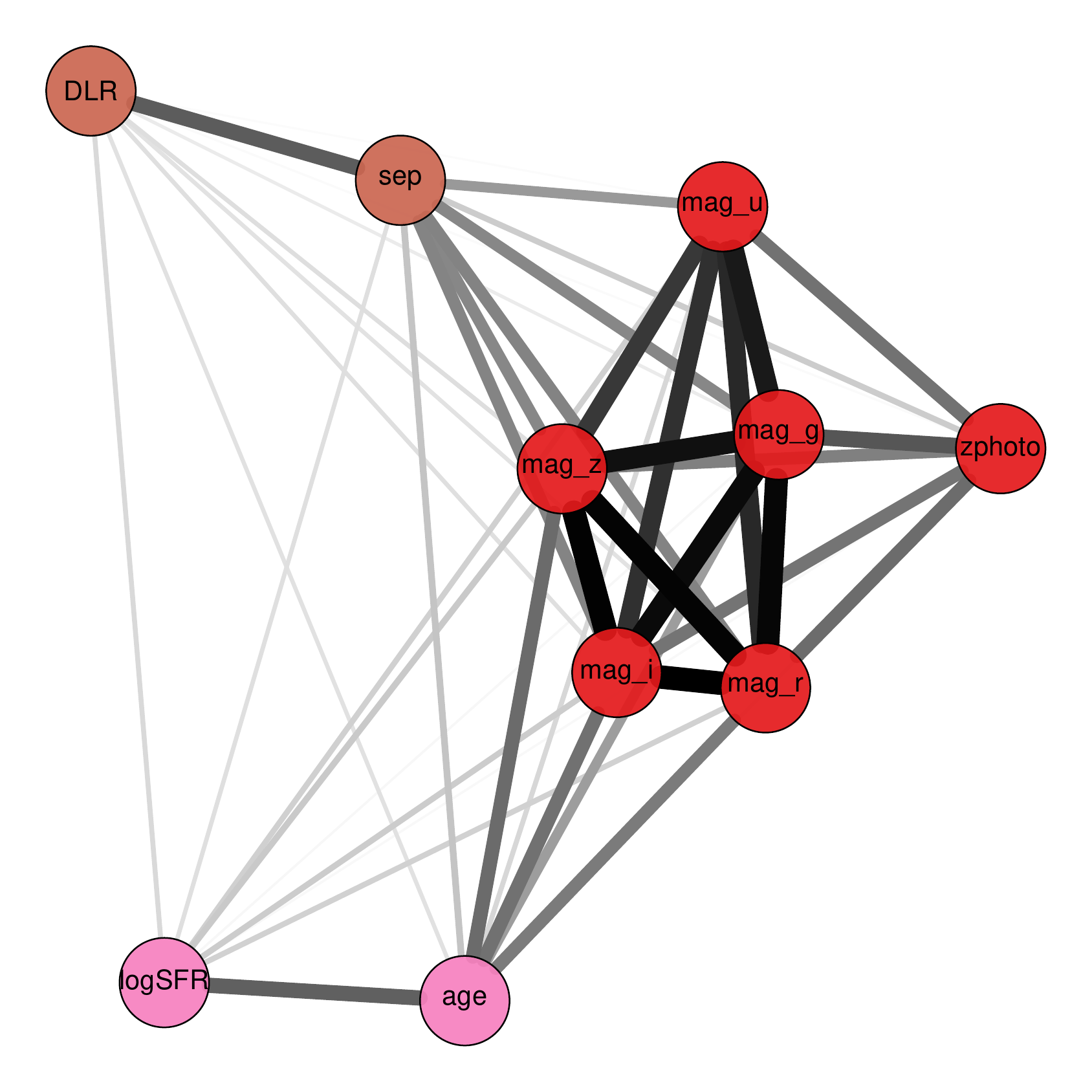}
\includegraphics[trim= 0.2cm 0.5cm 0.2cm 0.5cm, clip=true,width=1\columnwidth]{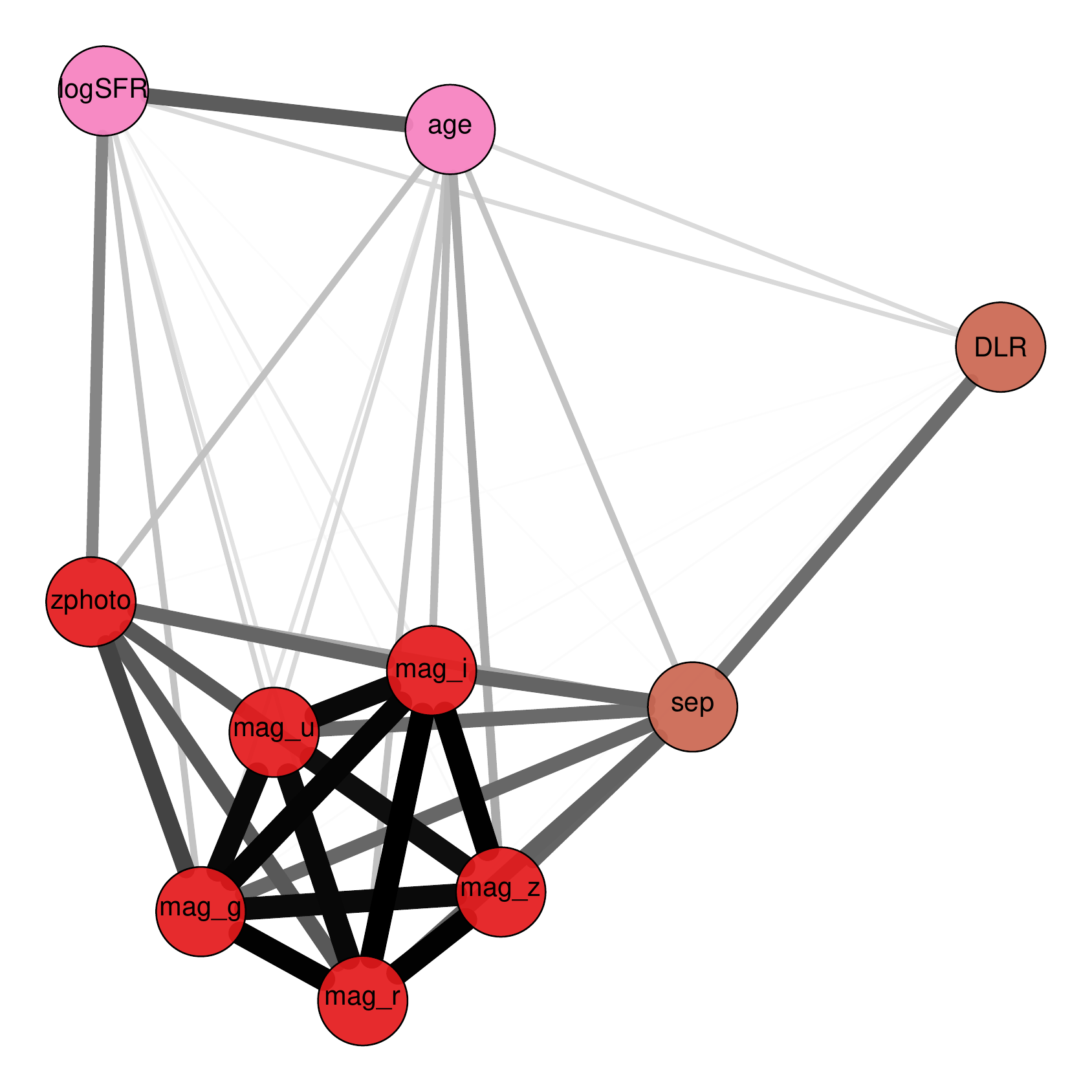}
\caption{ Graph representation   of the host  galaxy properties from \citet{Sako2014}.  The thickness of the edges are weighted by the degree of correlation between each pair of variables. 
The  width and color correspond to the degree of  association: the higher the correlation, the thicker and more color saturated the edge is. The left (right) side represents the properties of Type Ia (Type II) supernova host galaxies.} 
\label{fig:graph}
\end{figure*}

Graphs are powerful tools to  represent multivariate data and their relationships.  Examples of scientific applications are the analysis of cellular networks \citep{Aittokallio2006},  protein interactions \citep[e.g., Fig. 1 from][]{Aragues2006}, and brain disorders \citep[Fig. 2 from][]{Fornito2015}.
A graph  is defined by a set  of vertices  representing  the objects of study,  and a set of    edges  representing the relationships between them. There are many criteria for judging an optimally drawn 
graph such as:
 
\begin{itemize}

\item edge crossings should be minimized;
\item the vertices should be evenly distributed  in the plane;
\item the graph should reflect intrinsic symmetries; 
\item the edges should not cross nodes.
\end{itemize}
 Each item above can be understood as an  optimization problem, which is the subject of interest of a research field known as  {\it graph drawing} \citep[e.g.,][]{Tamassia2007}.
 There are several methods for graph representations. In this work we use the so-called {\it spring-embedder} algorithm \citep{eades84,Fruchterman91}. The  underlying idea is to allow  the vertices to behave like particles  moving  under the influence of repulsive and attractive  forces until the system reaches equilibrium.  This graph-drawing algorithm is particularly useful for graphs where the directions of the edges are not important, which is the case of a correlation matrix representation.   Fig. \ref{fig:graph} displays the correlations among properties of galaxies hosting Type Ia (left) and Type II (right) supernova. Each vertex represents a galaxy property, while  the thickness of the edges are weighted by the degree of correlation between each pair of variables \citep{Epskamp2012}. 
More specifically, the  width and color
of the edges correspond to the absolute value of the correlations: the higher the correlation,
the thicker and more saturated the edge is. 
Highly correlated parameters appear  closer in the graph.

%%%%%%%%%%%%%%%%%%%%%%%%%%%%%%%%%%%%%%%%%%%%%%%%%%%%%%%%%%%%%%%%%%%%%%%%%%%%%%%%%%%%%%%%%%%%%%%%%%%%%%%%%%%%%%%%%%%%%%%%%%%%%%%%%%%%

%%%%%%%%%%%%%%%%%%%%%%%%%%%%%%%%%%%%%%%%%%%%%%%%%%%%%%%%%%%%%%%%%%%%%%%%%%%%%%%%%%%%%%%%%%%%%%%%%%%%%%%%%%%%%%%%%%%%%%%%%%%%%%%%%%%%

\subsection{Chord diagram}

Chord diagram  is a flexible and popular tool that has been  used in many different applications, such as identification of  relevant signatures in  cancer genome  \citep[Fig. 1 from][]{Bunting2013}, or study of the relation between foragers and farmers  in Central Europe during the \textit{Stone Age} \citep[Fig. S5 from][]{Bollongino2013}.

\begin{figure*}
\centering
\includegraphics[width=1\columnwidth]{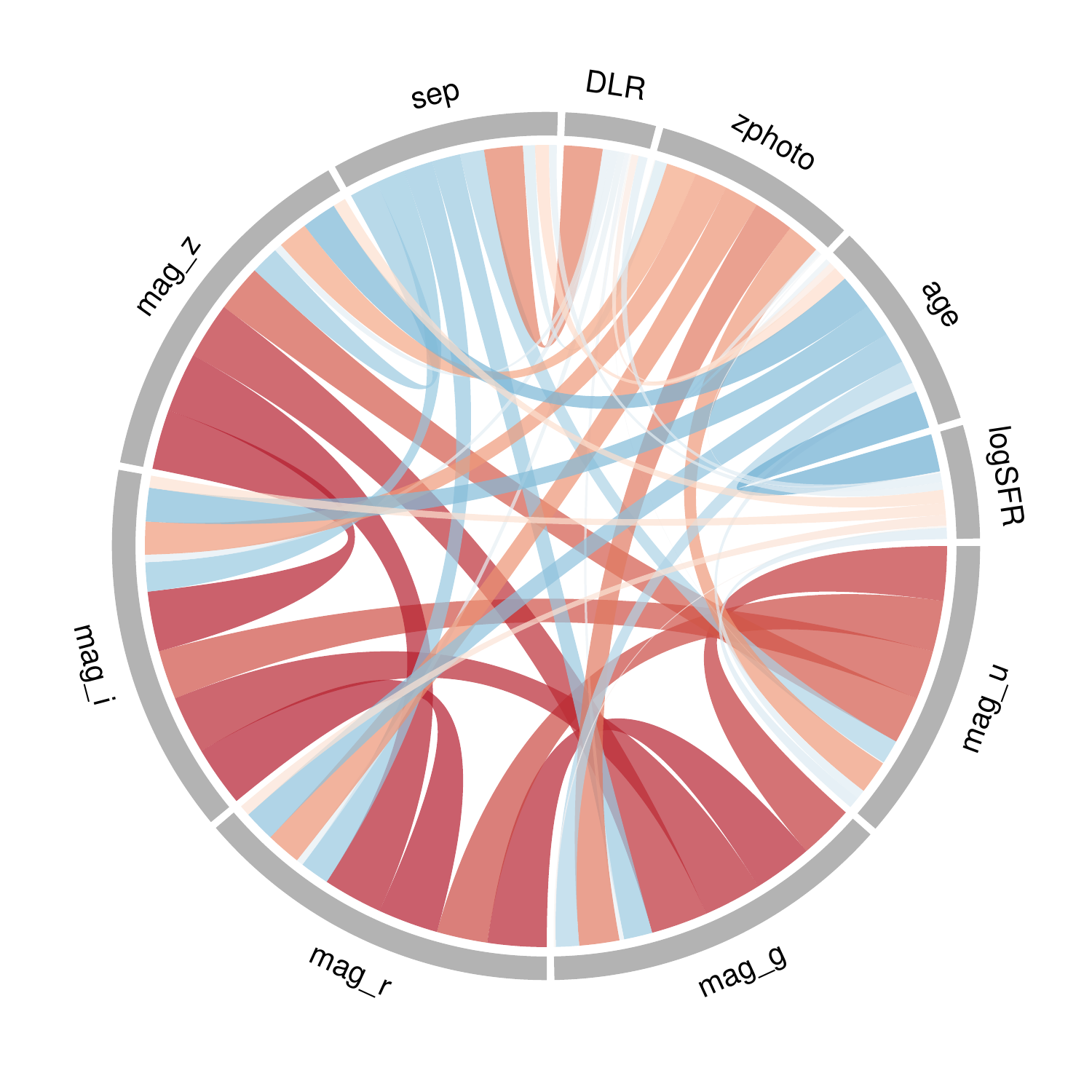}
\includegraphics[width=1\columnwidth]{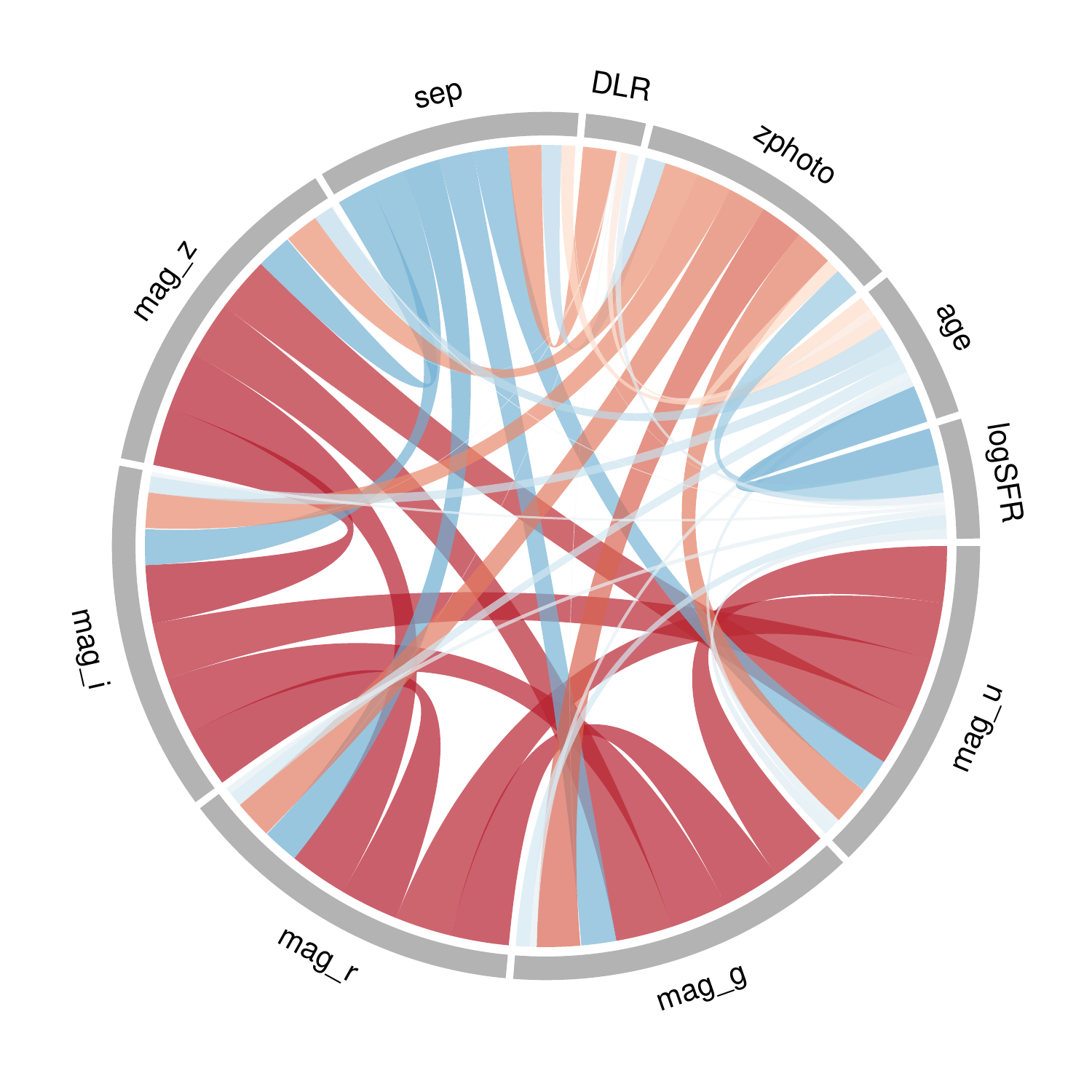}
\caption{A chord diagram representing the Pearson correlations among the galaxy properties hosting Type Ia (left panel), and Type II supernovae (right panel). 
  }
\label{fig:chord}
\end{figure*}

In the case studied here, the chord diagram represents another visualization of the correlation matrix, likewise the graph, heatmap and  dendrogram.  This tool illustrates relationships between distinct parameters. The columns and rows are represented by segments around the circle. Individual cells are shown as ribbons, which connect the corresponding row and column segments \citep{Gu2014}. The thickness of the ribbons is weighted by the degree of correlation between each pair of variables.  Fig. \ref{fig:chord}  portrays  the correlations among supernova Type Ia/II host galaxy properties. For a given choice of colour palette, the colour intensity ranges from fully anti-correlated to correlated values.

%%%%%%%%%%%%%%%%%%%%%%%%%%%%%%%%%%%%%%%%%%%%%%%%%%%%%%%%%%%%%%%%%%%%%%%%%%%%%%%%%%%%%%%%%%%%%%%%%%%%%%%%%%%%%%%%%%%%%%%%%%%%%%%%%%%%

%%%%%%%%%%%%%%%%%%%%%%%%%%%%%%%%%%%%%%%%%%%%%%%%%%%%%%%%%%%%%%%%%%%%%%%%%%%%%%%%%%%%%%%%%%%%%%%%%%%%%%%%%%%%%%%%%%%%%%%%%%%%%%%%%%%%

\subsection{Nightingale chart}

The last plot is inspired by the original  \textit{Nightingale  chart}  \citep[e.g.,][]{Cohen1984,McDonald2001}. 
This  is  one of the most influential statistical visualizations of all time,  used by Florence Nightingale to convince Queen Victoria about improving hygiene in military hospitals \citep[see also][for a review of radial methods in information visualization]{Draper2009}. 

\begin{figure*}
\centering
\includegraphics[width=2\columnwidth]{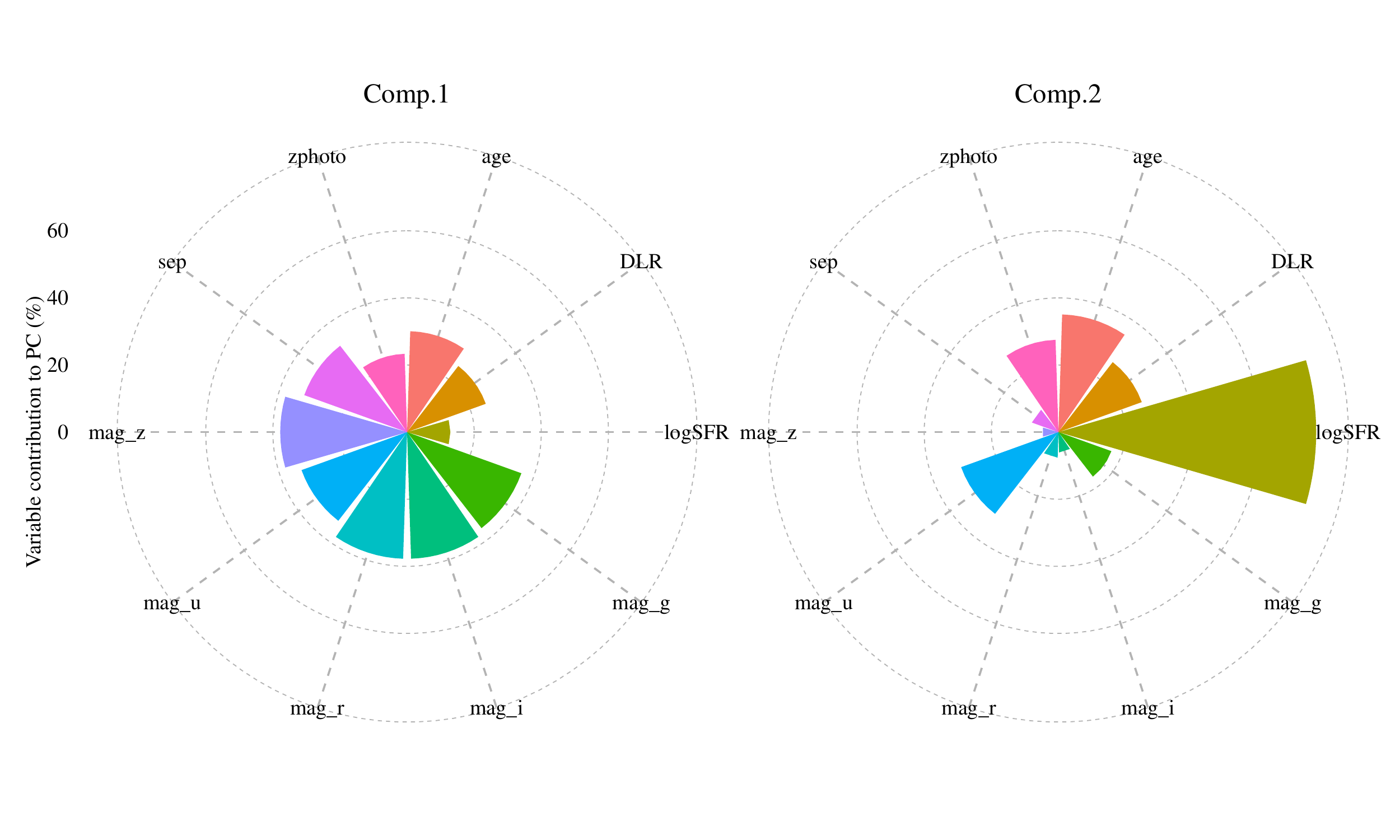}
\includegraphics[width=2\columnwidth]{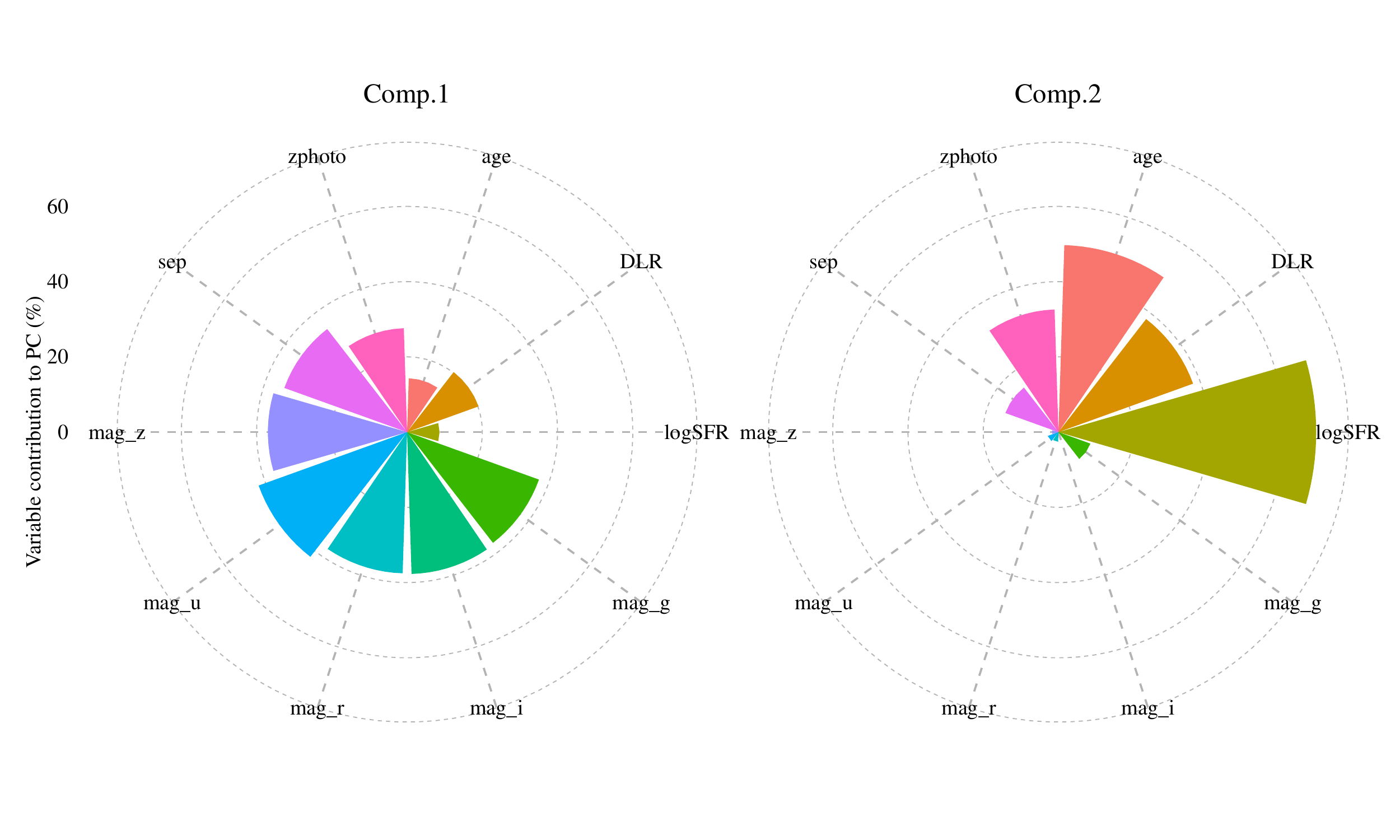}
\caption{A Nightingale diagram representing the contributions of the galaxy properties hosting Type Ia (left panel) and Type II (right panel) supernovae.} 
\label{fig:sunburst}
\end{figure*}

We show it as  a  polar bar plot, where  the length of each slice   represents the relative contribution of each variable to the $i$-th Principal Component.  Fig \ref{fig:sunburst} displays the contributions of the supernova Type Ia/II host galaxy properties for the first and second principal components\footnote{We should warn the reader that currently the {\sc shiny} interface does not work  well with  more than 4 PCs  simultaneously displayed on the screen. This limitation  can be potentially fixed by tweaking the  figure dimensions, if e.g. a {\sc pdf} file is produced using the {\sc r} command line (see \ref{app:shiny}).}.

%%%%%%%%%%%%%%%%%%%%%%%%%%%%%%%%%%%%%%%%%%%%%%%%%%%%%%%%%%%%%%%%%%%%%%%%%%%%%%%%%%%%%%%%%%%%%%%%%%%%%%%%%%%%%%%%%%%%%%%%%%%%%%%%%%%%

%%%%%%%%%%%%%%%%%%%%%%%%%%%%%%%%%%%%%%%%%%%%%%%%%%%%%%%%%%%%%%%%%%%%%%%%%%%%%%%%%%%%%%%%%%%%%%%%%%%%%%%%%%%%%%%%%%%%%%%%%%%%%%%%%%%%

%%%%%%%%%%%%%%%%%%%%%%%%%%%%%%%%%%%%%%%%%%%%%%%%%%%%%%%%%%%%%%%%%%%%%%%%%%%%%%%%%%%%%%%%%%%%%%%%%%%%%%%%%%%%%%%%%%%%%%%%%%%%%%%%%%%%

\section{Summary}

\label{sec_summary}

We have presented the AMADA package, a web application for interactive exploration and information retrieval of high-dimensional datasets.   This is designed for high-dimensional catalogs, with a wide range of applications. 
There are, though, some limitations in terms of data-size and performance. In particular, 
 {\sc shiny} allows to upload in the application only up to 1GB of data. Thus, the {\sc shiny} server  should be mostly used for a quick exploration of the package features, so that the user can skip the installation  step to familiarize with the code, while we recommend  to run AMADA  locally (as explained in \ref{app:shiny}) when applied to a real scientific problem.  
In addition, the speed performance of some methods, such as the hierarchical clustering, may not scale well  with  very  large  datasets. As a reference, the processing time to produce a dendrogram from a matrix with 100,000 objects and 100 columns was $\sim 1.5$ seconds on an iMac  featuring a 3,5 GHz Intel Core i7 and 32 GB of ram memory. An example of the script to reproduce this test is given below, 
\begin{lstlisting} 
require(AMADA)
N = 100000#Number of rows
M= 100# Number of columns
M1<-matrix(rnorm(N*M,mean=0,sd=1), N, M) 
ptm <- proc.time()
corr<-Corr_MIC(M1,"pearson")
Fig1<-plotdendrogram(corr,"fan")
proc.time() - ptm
\end{lstlisting}
Therefore, despite some limitations, we expect the current version of the package to be suitable for a wide variety of astronomical catalogs.
Since this is a software release paper, we avoided a detailed scientific discussion on the available datasets, which here have been used merely as  a proof of concept.  However, it is worth mentioning that AMADA automatically recovers  and displays trivial and non-trivial correlations.  An example of the former is the correlation between the u, g, r, z and i magnitudes  of supernova host galaxies as seen in Fig. \ref{fig:graph}, while an example of the latter is the  association  between the star formation rate and u-band magnitude in the galaxy mock  catalog as shown in Fig. \ref{fig:dendrogram}. It is important to mention that few methods herein implemented are  a later   development of a previous work from the authors making use of MIC statistics and  robust PCA 
to understand   the redshift dependence of halo baryonic properties in the early Universe \citep{deSouza2014}. We therefore refer the reader to this work  as an example of application in a cosmological context of  the methods discussed here.

The code is freely available on {\sc github} and can be run both online and locally. This work is part of a larger enterprise known as  \textit{Cosmostatistics Initiative} (COIN)\footnote{\url{http://goo.gl/rQZSAB}}, whose philosophy is to  enable  astronomers to easily introduce novel  techniques into their daily research.
This  is an  open-source project, and  we expect to continuously add extra  features. Therefore,  we encourage the users to contact the authors with suggestions, while  potential contributors and developers can fork the AMADA repository on {\sc github}\footnote{\url{https://github.com/COINtoolbox/AMADA}}.

\section*{Acknowledgements}
We thank E. E. O. Ishida  for   the careful review and fruitful comments of  the manuscript. We thank M. L. Dantas and T. Kitching for  testing AMADA  on their respective machines.  We thank the constructive suggestions of the referee. 
The IAA Cosmostatistics Initiative (COIN)\footnote{\url{https://asaip.psu.edu/organizations/iaa/iaa-working-group-of-cosmostatistics}} is a non-profit organization whose aim is to nourish the synergy between astrophysics, cosmology, statistics and machine learning communities.

%%%%%%%%%%%%%%%%%%%%%%%%%%%%%%%%%%%%%%%%%%%%%%%%%%%%%%%%%%%%%%%%%%%%%%%%%%%%%%%%
%BIBLIOGRAPHY
%%%%%%%%%%%%%%%%%%%%%%%%%%%%%%%%%%%%%%%%%%%%%%%%%%%%%%%%%%%%%%%%%%%%%%%%%%%%%%%%

 % /--------------------
% APPENDIX

\appendix
\section{Running AMADA locally}
\label{app:shiny}

\subsection{From Shiny}

To install and run the interface, the first step is to have \textsc{r} in your computer\footnote{\url{http://www.r-project.org}}.  Thereafter, you have to install the following \textsc{r} packages:

\begin{lstlisting} 
install.packages(c("ape","circlize","corrplot","devtools","fpc","ggplot2","ggthemes","MASS","markdown","mclust","minerva","mvtnorm","pcaPP","pheatmap","phytools","qgraph","RColorBrewer","RCurl","squash","stats","shiny"),dependencies=TRUE)
\end{lstlisting}
We are now read to install AMADA from GitHub repository:
\begin{lstlisting} 
require(devtools)
install_github("RafaelSdeSouza/AMADA")
\end{lstlisting}
An alternative simpler option  is to type the following command 
\begin{lstlisting} 
require(devtools)
install_github("COINtoolbox/AMADA",dependencies=TRUE)
\end{lstlisting}
and \textsc{r} will automatically install the necessary dependencies to run AMADA.
After installing the AMADA package, the user can run the visual interface with the following command: 
  
 \begin{lstlisting} 
require(shiny)
runUrl("https://github.com/COINtoolbox/AMADA_shiny/archive/master.zip")
\end{lstlisting}

AMADA  can also  be used directly  via the web. This option requires no local installation, but the actual processing may be slower. This web interface is hosted by the shinyapps.io platform\footnote{\url{http://www.shinyapps.io}}, and can be accessed directly at \url{http://goo.gl/UTnU7I}.

\subsection{From R command line}

If the user prefer to run AMADA on its own data   without relying on the shiny interface, it can be done  directly from {\sc r} command line. An example of how to produce a dendrogram of the Type Ia supernova dataset and saving it as a {\sc pdf} file is presented below: 
\begin{lstlisting} 
require(AMADA) #Load the package
data("SNIa") #Load the SNIa data
corr<-Corr_MIC(SNIa,"pearson")
Fig1<-plotdendrogram(corr,"phylogram")
\end{lstlisting}
To save the figure as {\sc pdf} file,  with a customized height and width, just type the following:
\begin{lstlisting}
pdf("phylogram.pdf",height = 8,width=8)
Fig1
dev.off()
\end{lstlisting}
Examples of how the use the other functions inside {\sc r} can be found in the description file, which can be access via the command\footnote{We should stress that the functions to display the chord diagram and the heatmap are basically convenient  wrappers to the functions available in the packages {\sc pheatmap}  and {\sc circlize}.} 
\begin{lstlisting}
help(package="AMADA") 
\end{lstlisting}
In the current package version, the layout of the figures is mostly hardcoded, but it can be easily changed inside the source code. We expect to add more flexibility in future versions.

\end{document}